\documentclass{bvm} 

\begin{document}

\newcommand{\bvmyear}{2024}
\newcommand{\mymathrm}[1]{\operatorname{\textrm{\spaceskip.15em#1}}}

\selectlanguage{english} 

\title{Automated Volume Corrected Mitotic Index Calculation Through Annotation-Free Deep Learning using Immunohistochemistry as Reference Standard}


\titlerunning{Automated Volume Corrected Mitotic Index}

\authorrunning{}

\author{
	Jonas \lname{Ammeling} \inst{1}, 
    Moritz \lname{Hecker} \inst{1},
    Jonathan \lname{Ganz} \inst{1}, 
    Taryn A. \lname{Donovan} \inst{2},
    Robert \lname{Klopfleisch} \inst{3},
    Christof A. \lname{Bertram} \inst{4},
	Katharina \lname{Breininger} \inst{5},
	Marc \lname{Aubreville} \inst{1}
}
\authorrunning{Ammeling et al.}

\institute{
\inst{1} Technische Hochschule Ingolstadt, Ingolstadt, Germany\\
\inst{2} The Schwarzman Animal Medical Center, New York, USA \\
\inst{3} Institute of Veterinary Pathology, Freie Universität Berlin, Germany \\
\inst{4} Institute of Pathology, University of Veterinary Medicine Vienna, Vienna, Austria\\
\inst{5} Department Artificial Intelligence in Biomedical Engineering, Friedrich-Alexander-Universität Erlangen-Nürnberg, Erlangen, Germany\\
}

\email{jonas.ammeling@thi.de}

\maketitle

\begin{abstract}
The volume-corrected mitotic index (M/V-Index) was shown to provide prognostic value in invasive breast carcinomas. However, despite its prognostic significance, it is not established as the standard method for assessing aggressive biological behaviour, due to the high additional workload associated with determining the epithelial proportion. In this work, we show that using a deep learning pipeline solely trained with an annotation-free, immunohistochemistry-based approach, provides accurate estimations of epithelial segmentation in canine breast carcinomas. We compare our automatic framework with the manually annotated M/V-Index in a study with three board-certified pathologists. Our results indicate that the deep learning-based pipeline shows expert-level performance, while providing time efficiency and reproducibility.  
\end{abstract}

\section{Introduction}

The accurate assessment of mitotic activity in histopathology plays an essential role in cancer diagnosis, prognosis, and treatment decisions. The mitotic count (MC), which represents the number of mitotic figures in a given area of tissue is a key parameter in many grading schemes used to assess the proliferation rate and aggressiveness of various malignancies. The prognostic significance of the MC is limited due to high inter-observer variability, poor reproducibility and the labor-intensive nature of this microscopic task. Additionally, varying cellular densities of different tumors can limit the interpretability of the MC across cases. Haapasalo \& Collan \cite{Haapasalo1989} introduced the volume corrected mitotic index (M/V-Index) in an effort to standardize the counting of mitotic figures. The M/V-Index standardizes the MC by dividing it by the area fraction of the epithelial tissue estimated subjectively or by using a point grid and adjusting it for the size of a high power field, resulting in an estimate of the number of mitotic figures per square millimeter. Jannink et al. \cite{Jannink1995} demonstrated that the M/V-Index, along with tumor size and lymph node status, offered better prognostic information for human breast cancer than the uncorrected MC. Despite its significance, the M/V-Index is not widely adopted due to the additional effort required in estimating the epithelial tissue fraction, leading to higher inter-observer variability. Hence, the faster and simpler uncorrected MC remains the preferred method for assessing tumor proliferation. In this study, we provide an automated framework for the calculation of the M/V-Index on hematoxilin and eosin (H\&E)-stained images. The framework is developed in an annotation-free fashion (i.e. not requiring any human labelling effort for estimating the area fraction of the epithelial tissue) by using immunohistochemistry (IHC) as a reference standard \cite{Bulten2019} and leveraging an existing model for MC estimation \cite{Wilm2022}. This framework provides the first proof of concept that the M/V-Index can be estimated with accuracy, efficiency and reproducibility, offering a more objective method to assess tumor proliferation.

\section{Materials}

The dataset consisted of 50 canine mammary carcinoma samples collected at the University of Veterinary Medicine, Vienna. 
The samples were first stained with standard H\&E and scanned with a 3DHistech Panoramic Scann II at 40$\times$ magnification (0.25 \textmu m/px). After scanning, the slides were destained and then restained with the pan-cytokeratin AE1/AE3 primary antibody, which is specific for cytokeratin proteins commonly found in epithelial tissues. The IHC slides were rescanned using the same scanner and magnification. The process of restaining the slides resulted in 50 H\&E and IHC whole slide image (WSI) pairs which were co-registered using a robust quad-tree based WSI registration method \cite{Marzahl2021registration}. Due to some staining artefacts and alignment errors, 9 samples were removed from the dataset. Of the remaining 41 samples, 12 were kept as a hold-out test set. The remaining 29 samples were used in a 5-fold Monte Carlo cross-validation where the samples were randomly divided into 20 training and 9 validation cases. For each slide in the hold-out test set, a region of interest (ROI) with an area of $2.37\ts \mathrm{mm^2}$ equivalent to 10 high power fields (HPFs) in a microscope with an ocular Field Number (FN) of $22\ts\mathrm{mm}$ was selected and annotated for epithelial tissue by a board-certified pathologist.

\section{Methods}

To reduce the overall amount of manual labelling for our automated M/V-Index system, we automatically generated the training data for tumor epithelium segmentation from the IHC slides, which were then transferred by the registration method to the H\&E slides on which we trained our segmentation network.


\subsection{Automatic Tumor Epithelium Segmentation on H\&E}

\begin{figure}[hbt]
    \centering
    \includegraphics[width=\textwidth]{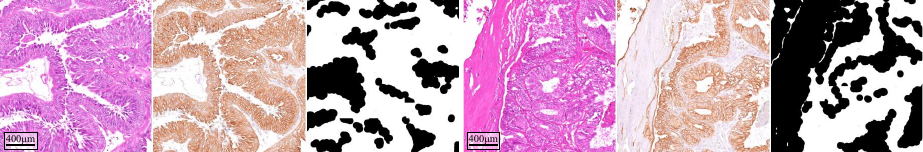}
    \caption{Two example patches from the mask generation process. (left) H\&E patch, (middle) IHC patch, (right) automatically generated mask (white: epithelial tissue, black: background).}
    \label{fig:train_set_masks}
\end{figure}

For tumor epithelium mask generation, we created an IHC map from downsampled WSIs, excluding irrelevant areas. The IHC map was created by applying color deconvolution and then using the cytokeratin channel to generate a binary mask by first applying a Gaussian blur filter and then Otsu's adaptive thresholding method, followed by a closing operation to remove small interruptions. Patches with at least 5\% non-zero values in the IHC map were then processed at full resolution, first using color deconvolution, followed by binary thresholding and an opening operation to remove small noise from staining artefacts and intensity variations. Finally, a closing operation with a large circular kernel resulted in a coarse segmentation mask (Fig. \ref{fig:train_set_masks}) that was helpful in mitigating small alignment errors from the registration process. Bulten et al. \cite{Bulten2019} used a similar pipeline on prostate samples, however they used additional human labelling to further optimize the masks. The tumor epithelium segmentation network was based on a U-Net architecture, consisting of an EfficientNet-b0 encoder and a classical encoder composed of up-sampling and convolutional layer. The network was trained on patches of size $1024 \times 1024$ at a resolution of 0.5 \textmu m/px and a batch size of $4$. We used the Adam optimizer with an initial learning rate of $0.001$ and an exponentially decaying learning rate schedule with a factor of $0.99$. The loss function consisted of a weighted combination with factor $0.5$ of dice loss and binary cross-entropy loss. To make the model more robust to the noisy training masks, we used label smoothing with a factor of $0.1$ for the targets in the cross-entropy loss function. The model was trained with standard online augmentation including random flipping, rotation, Gaussian blurring, and changes in brightness, contrast, saturation and hue.

\subsection{Mitotic Count Estimation}

The mitotic count estimation network was based on DA-RetinaNet \cite{Wilm2022}, a one-stage RetinaNet object detector with a ResNet-18 backbone, trained with domain adversarial training on the MIDOG 21 \cite{Aubreville2023} training dataset, which consists of 200 human breast cancer images from four different scanners. No further annotations or fine tuning were considered necessary due to the morphological similarity between the two species. For further details on the implementation of the network, the reader is referred to Wilm et al.~\cite{Wilm2022}.

\subsection{Volume Corrected Mitotic Index}

The volume corrected mitotic index (M/V-Index) was originally proposed by Haapasalo \& Collan \cite{Haapasalo1989} in order to standardize the MC based on the cellular density of the tumor. The formula for calculating the M/V-Index is 

\begin{equation*}
        \mymathrm{M/V-Index} = k \sum_{i=1}^n \frac{\mathrm{MC_i}}{\mathrm{Vv_i}}, \mathrm{where}
\end{equation*}

\noindent $n$ is the number of microscope fields studied, MC is the number of mitotic figures in a selected field, Vv is the volume fraction (in per cent) of the neoplastic tissue in the same field, either estimated subjectively or with point-counting, and $k$ is a coefficient characterizing the microscope: $k = 100 / \pi r^2$ where $r$ in (in $\mathrm{mm}$) is the radius of the circular microscope field. To adapt the formula for a digital evaluation we defined $k=100/A$, where $A$ is the area (in $\mathrm{mm^2}$) of the evaluated ROI. Here we set $k=100/2.37\ts \mathrm{mm^2}$, where $2.37~\mathrm{mm^2}$ is the area of 10 HPFs at 40$\times$ magnification (0.25 \textmu m/px). The MC was estimated by first dividing the image into overlapping patches of size $512 \times 512$ at 40$\times$ magnification (0.25 \textmu m/px). The predictions of the MC estimation model are then fused and transformed into the original coordinate space. Similarly, Vv is estimated using the segmentation model on overlapping patches of size $1024 \times 1024$ at  20$\times$ magnification (0.5 \textmu m/px). The concatenated result of the epithelium segmentation is then used as a mask to filter mitotic figures that are within the epithelial tissue region. Finally, the M/V-Index is calculated as $k \times \mathrm{MC} / \mathrm{Vv}$ over the entire ROI.
 
 \begin{figure}[hbt]
    \centering
    \includegraphics[width=\textwidth]{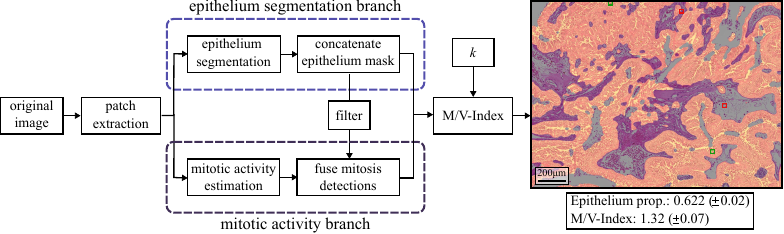}
    \caption{Automated volume corrected mitotic index (M/V-Index) framework. The image on the right shows the epithelium segmentation in orange. Green boxes represent mitotic figures detected within the epithelium mask. Red boxes represent mitotic figures filtered out by the epithelium mask.  The results below the image present algorithm-derived mean and standard deviation for both epithelium proportion and the M/V-Index.}
    \label{fig:main}
\end{figure}

\subsection{Manual vs. Automatic M/V-Index}

Three board-certified pathologists manually determined the M/V-Index using a Weibel point grid \cite{Jannink1995} to reduce interrater variability. The grid, adjusted to the ROI size, had 432 points, equivalent to the 42-point Weibel grid used for a single HPF. Pathologists separately annotated epithelium proportion and mitotic figures, measuring the time for each task. We compare the estimated epithelium proportion using the Weibel grid to the ground truth calculated from the epithelium masks of our pathologist. The ground truth for the M/V-Index is calculated by first filtering all mitotic figure annotations from our pathologists through the ground truth epithelium mask and then averaging the M/V-Index of the individual pathologists.

\section{Results}

\begin{figure}[bht]
    \centering
    \includegraphics[width=\textwidth]{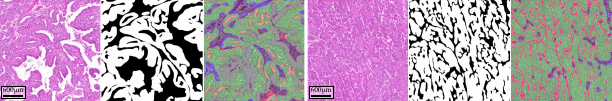}
    \caption{Tumor epithelium segmentation results on the hold-out test set. (left) original image, (middle) expert labelled ground truth, (right) segmentation results. Green pixels show true positives, red false positives and blue false negatives.}
    \label{fig:qualitativ_results}
\end{figure}

\begin{figure}[bht]
    \centering
    \includegraphics[width=\textwidth]{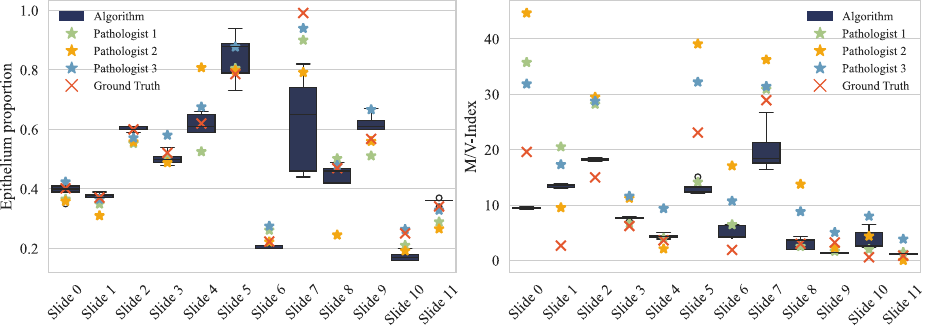}
    \caption{Results of the manual vs. automatic M/V-Index annotation study with three board-certified pathologist. (left) Epithelium proportion, (right) M/V-Index.}
    \label{fig:study_results}
\end{figure}

The epithelium segmentation algorithm achieved an average intersection-over-union (IOU) of $0.71~(\pm 0.01)$ and F1 score of $0.83~(\pm 0.01)$ on the hold-out test set. Qualitative results are in Fig. \ref{fig:qualitativ_results}. A comparison of results between manual and automatic M/V-Index is in Tab. \ref{tab:study_results}. The mean absolute error (MAE) for epithelium proportion was consistent among pathologists and the algorithm, shown in Fig. \ref{fig:study_results} (left). For the M/V-Index, the average MAE was $7.39~(\pm 1.37)$ for pathologists and $4.51~(\pm 0.25)$ for the algorithm, displayed in Fig. \ref{fig:study_results} (right). In our study, pathologists took an average of 12 minutes, while the algorithm only took 20 seconds to calculate the M/V-Index on a ROI. Using the Weibel grid, pathologists spent an average of 6 minutes assessing only the epithelium proportion per ROI.

\begin{table}[thb]
\centering
\setlength{\tabcolsep}{6pt}
\caption{Results from the comparison between the manual and automatic M/V-Index to the ground truth. Displayed are the mean absolute error (MAE), and Pearson's correlation coefficient.}
\begin{tabular}{lll} 
\hline
                                  & MAE           & Pearson's r \\ \hline
\multicolumn{3}{l}{\textit{Epithelium Proportion}}                                     \\ 
Pathologists                      & 0.06 $\pm$ 0.02 & 0.95 $\pm$ 0.04   \\ 
\multicolumn{1}{l}{Algorithm}     & 0.06 $\pm$ 0.01 & 0.83 $\pm$ 0.11   \\ \hline
\multicolumn{3}{l}{\textit{M/V-Index}}                                                      \\ 
Pathologists                     & 7.39 $\pm$ 1.37 & 0.87 $\pm$ 0.06   \\
Algorithm                        & 4.51 $\pm$ 0.25 & 0.78 $\pm$ 0.02   \\  \hline
\end{tabular}%
\label{tab:study_results}
\end{table}

\section{Discussion}


We showed that our algorithm can estimate the M/V-Index comparable to expert level performance. We avoided the need for time-consuming manual labeling of tumor tissue by the innovative use of immunohistochemical stainings as the ground truth. The evaluation of the model on a expert-derived ground truth of the test set serves as a first proof of concept that the M/V-Index can be calculated in an automated, fast and reproducible way. This framework is easily extendable to calculate the M/V-Index on entire WSIs automatically providing even larger time savings for the pathologist and further reducing the potential for inter-observer variability by selecting the region of highest mitotic activity in a computer-aided and more reproducible fashion. In a future work, this framework can be improved by employing a more precise IHC marker during training, addressing the issue caused by pan-cytokeratin also binding to myoepithelial cells, which is a potential error for the segmentation model.


\begin{acknowledgement}
   This work was supported by the Bavarian Institute for Digital Transformation (bidt) under the grant ``Responsibility Gaps in Human Machine Interaction (ReGInA)''
\end{acknowledgement}

\printbibliography

\end{document}